\documentclass[twocolumn,pra,floatfix,showpacs,superscriptaddress]{revtex4}
\usepackage{amssymb}
\usepackage{amsmath}
\usepackage{graphicx}
\usepackage{float}

\def\be{\begin{equation}}
\def\ee{\end{equation}}
\def\omegar{\omega_{\rm r}}
\def\omegad{\omega_{\mu \rm w}}

\def\ndet{n_{\rm amp}}
\def\ncrit{n_{\rm crit}}
\newcommand{\eq}[1]{(\ref{#1})}
\newcommand{\id}{\openone}
\newcommand{\ket}[1]{\left|#1\right>}
\newcommand{\bra}[1]{\left< #1 \right|}
\def\ua{\uparrow}
\def\da{\downarrow}

\begin{document}

\title{Cavity quantum electrodynamics for superconducting electrical
circuits: an architecture for quantum computation}
\author{Alexandre Blais}
\affiliation{Departments of Physics and Applied Physics, Yale
University, New Haven, CT 06520}
\author{Ren-Shou Huang}
\affiliation{Departments of Physics and
Applied Physics, Yale University, New Haven, CT 06520}
\affiliation{Department of Physics, Indiana University,
Bloomington, IN 47405}
\author{Andreas Wallraff}
\affiliation{Departments of Physics and Applied Physics, Yale
University, New Haven, CT 06520}
\author{S. M. Girvin}
\affiliation{Departments of Physics and Applied Physics, Yale
University, New Haven, CT 06520}
\author{R. J. Schoelkopf}
\affiliation{Departments of Physics and Applied Physics, Yale
University, New Haven, CT 06520}
\date{\today}
%, 42.50.-p}

\begin{abstract}
We propose a realizable architecture using one-dimensional
transmission line resonators to reach the strong coupling limit of
cavity quantum electrodynamics in superconducting electrical
circuits. The vacuum Rabi frequency for the coupling of cavity
photons to quantized excitations of an adjacent electrical circuit
(qubit) can easily exceed the damping rates of both the cavity and
the qubit.  This architecture is attractive both as a macroscopic analog
of atomic physics experiments and for quantum computing
and control, since it provides strong inhibition of spontaneous
emission, potentially leading to greatly enhanced qubit lifetimes,
allows high-fidelity quantum non-demolition measurements of the
state of multiple qubits, and has a natural mechanism for
entanglement of qubits separated by centimeter distances. In
addition it would allow production of microwave photon states of
fundamental importance for quantum communication.
\end{abstract}

\maketitle

%
%%%%%%%%%%%%%%%%%%%%%%%%%%%
% I: Intro and cQED Review
%%%%%%%%%%%%%%%%%%%%%%%%%%%

\section{Introduction}

Cavity quantum electrodynamics (cQED) studies the properties of
atoms coupled to discrete photon modes in high $Q$ cavities.  Such
systems are of great interest in the study of the fundamental quantum
mechanics of open systems, the engineering of quantum states and
the study of measurement-induced decoherence
\cite{mabuchi:2002,hood:2002,raimond:2001}, and have also been
proposed as possible candidates for use in quantum information
processing and transmission
\cite{mabuchi:2002,hood:2002,raimond:2001}. Ideas for novel cQED
analogs using nano-mechanical resonators have recently been
suggested by Schwab and collaborators
\cite{armour:2002,elinor:2003}. We present here a realistic proposal
for cQED via Cooper pair boxes coupled to a one-dimensional (1D)
transmission line resonator, within a simple circuit that can be
fabricated on a single microelectronic chip. As we discuss, 1D
cavities offer a number of practical advantages in reaching the
strong coupling limit of cQED over previous proposals using
discrete LC circuits \cite{makhlin:2001,buisson:2001}, large
Josephson junctions
\cite{marquandt:2001,plastina:2003,blais:2003}, or 3D cavities
\cite{al-saidi:2001,yang:2003,younoria:2003}.  Besides the
potential for entangling qubits to realize two-qubit gates
addressed in those works, in the present work we show that the cQED
approach also gives strong and controllable isolation of the qubits from the
electromagnetic environment, permits high fidelity quantum
non-demolition (QND) readout of multiple qubits, and can produce
states of microwave photon fields suitable for quantum
communication. The proposed circuits therefore provide a simple
and efficient architecture for solid-state quantum computation, in
addition to opening up a new avenue for the study of entanglement
and quantum measurement physics with macroscopic objects. 
We will frame our discussion in a way that makes contact between the
language of atomic physics and that of electrical engineering.  

We begin in Sec.~\ref{sec:review_cQED}  with a brief general overview of cQED
before turning to a discussion of our proposed solid-state
realization of cavity QED in Sec.~\ref{sec:scQED}.
We then discuss in Sec.~\ref{sec:zero_detuning}
the case where the cavity and the qubit are tuned in resonance and
in Sec.~\ref{sec:large_detuning} the case of large detuning which
leads to lifetime enhancement of the qubit.  In Sec.~\ref{sec:read-out},
a quantum non-demolition read-out protocol is presented.
Realization of one-qubit logical operations is discussed
in Sec.~\ref{sec:one_qubit} and two-qubit entanglement in
Sec.~\ref{sec:many_qubits}.  We show in Sec.~\ref{sec:dfs} how to
take advantage of encoded universality and decoherence-free
subspace in this system.

\section{Brief review of cavity QED}
\label{sec:review_cQED}

\begin{figure}[t]
\begin{center}
\includegraphics[width=3.45in]{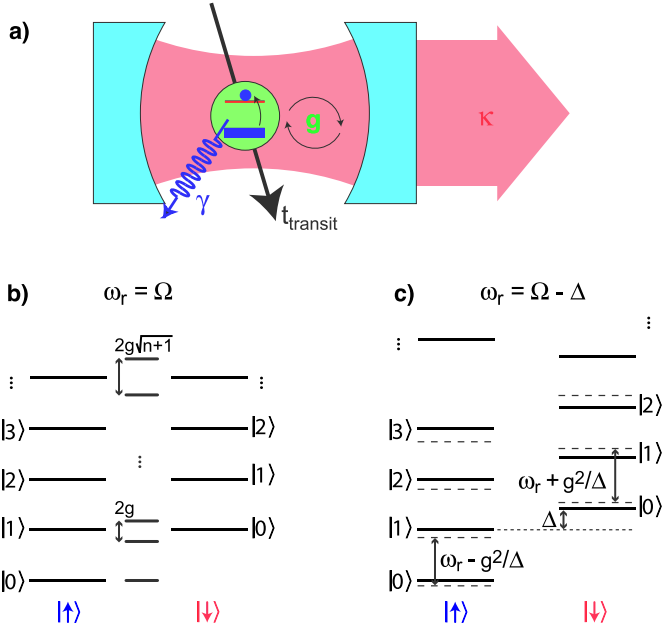} 
\caption{(color online). a) Standard representation of
cavity quantum electrodynamic system, comprising a single mode of
the electromagnetic field in a cavity with decay rate $\kappa$
coupled with a coupling strength $g={\cal E}_{\rm rms}d/\hbar$ to
a two-level system with spontaneous decay rate $\gamma$ and cavity
transit time $t_{\rm transit}$. 
b) Energy spectrum of the uncoupled (left and right) and dressed (center)
atom-photon states in the case of zero detuning.  The degeneracy of
the two-dimensional manifolds of states with $n-1$ quanta is lifted
by $2g\sqrt{n+1}$.
c) Energy spectrum in the dispersive regime (long dash lines).  To second order
in $g$, the level separation is independent of $n$, but depends on the state
of the atom.
}
\label{fig:cQED}
\end{center}
\end{figure}

Cavity QED studies the interaction between atoms and the quantized
electromagnetic modes inside a cavity.
In the optical version of cQED \cite{hood:2002}, schematically 
shown in Fig.~\ref{fig:cQED}(a), one drives the
cavity with a laser and monitors changes in the cavity
transmission resulting from coupling to atoms falling through the
cavity.  One can also monitor the spontaneous emission of the
atoms into transverse modes not confined by the cavity.  It is not
generally possible to directly determine the state of the atoms
after they have passed through the cavity because the spontaneous
emission lifetime is on the scale of nanoseconds. One can,
however, infer information about the state of the atoms inside the
cavity from real-time monitoring of the cavity optical
transmission.

In the microwave version of cQED \cite{raimond:2001}, one uses a
very high $Q$  superconducting 3D resonator to couple photons to
transitions in Rydberg atoms.  Here one does not directly monitor
the state of the photons, but is able to determine with high
efficiency the state of the atoms after they have passed through
the cavity (since the excited state lifetime is of order 30 ms).
{From} this state-selective detection one can infer information
about the state of the photons in the cavity.

The key parameters describing a cQED system (see Table~\ref{table:parameters})
are the cavity resonance frequency $\omegar$, the atomic transition
frequency $\Omega$, and the strength of the atom-photon coupling
$g$ appearing in the Jaynes-Cummings Hamiltonian \cite{walls-milburn}
\be
H = \hbar\omegar \left(a^\dagger a +
\frac{1}{2}\right) + \frac{\hbar\Omega}{2}\sigma^z + \hbar
g(a^\dagger\sigma^-+a\sigma^+) + H_\kappa+H_{\gamma}.
\label{eq:Jaynes-Cummings}
\ee
Here $H_\kappa$ describes the
coupling of the cavity to the continuum which produces the cavity
decay rate $\kappa = \omegar/Q$, while $H_{\gamma}$ describes the
coupling of the atom to modes other than the cavity mode which
cause the excited state to decay at rate $\gamma$ (and possibly
also produce additional dephasing effects). An additional
important parameter in the atomic case is the transit time
$t_{\rm{transit}}$ of the atom through the cavity.

%%%%%%%%%%%%%%%%%%%%%%%%%%%%%%%%%%%%%%%%%%%
\begin{table*}[ht]
%\begin{center}
\begin{small}
\hspace*{-1.5cm}\begin{tabular}{|l|l|c|c|c|}\hline
parameter&symbol&3D optical&3D microwave&1D circuit\\ \hline %
resonance/transition frequency&$\omega_{\rm{r}}/ 2\pi$, $\Omega/ 2 \pi$&$350 \, \rm{THz}$&$51 \, \rm{GHz}$& $10 \, \rm{GHz}$\\ \hline %
vacuum Rabi frequency&$g/\pi$, $g/\omegar$&$220 \, \rm{MHz}$, $3 \times 10^{-7}$&$47 \, \rm{kHz}$, $1 \times 10^{-7}$&$100 \, \rm{MHz}$, $5\times 10^{-3}$\\ \hline %
transition dipole&$d/e a_0$&$\sim 1$&$1 \times 10^3$&$ 2\times 10^4$\\ \hline%
cavity lifetime&$1/\kappa, Q$&$10 \, \rm{ns}$, $3 \times 10^7$ & $1 \, \rm{ms}$, $3 \times10^{8}$&$160 \, \rm{ns}$, $10^4$\\ \hline %
atom lifetime &$1/\gamma$&$ 61\,\rm{ns}$&$30 \, \rm{ms}$&$ 2 \, \rm{\mu s}$\\ \hline %
atom transit time&$t_{\rm transit}$&$\ge 50 \, \rm{\mu s}$&$100 \, \rm{\mu s}$&$\infty$\\ \hline %
critical atom number&$N_0=2\gamma\kappa/g^2$&$6 \times 10^{-3}$&$3\times10^{-6}$&$\leq 6 \times 10^{-5}$\\ \hline %
critical photon number&$m_0=\gamma^2/2g^2$&$3\times10^{-4}$&$3\times10^{-8}$&$\leq 1 \times 10^{-6}$\\ \hline %
\# of vacuum Rabi flops&$n_{\rm Rabi}= 2g/(\kappa+\gamma)$ &$\sim 10$& $\sim 5$ &$\sim 10^2$ \\ \hline %
\end{tabular}
\end{small}
%\end{center}
\caption{Key rates and cQED parameters for optical
\cite{hood:2002} and microwave \cite{raimond:2001} atomic systems
using 3D cavities, compared against the proposed approach using
superconducting circuits, showing the possibility for attaining
the strong cavity QED limit ($n_{\rm Rabi} \gg 1$). For the 1D
superconducting system, a full-wave ($L=\lambda$) resonator,
$\omegar/2\pi=10$ GHz, a relatively low $Q$ of $10^4$ and coupling
$\beta=C_g/C_\Sigma=0.1$ are assumed. For the 3D microwave case,
the number of Rabi flops is limited by the transit time. For the
1D circuit case, the intrinsic Cooper-pair box decay rate is
unknown; a conservative value equal to the current experimental
upper bound $\gamma \le 1/(2 \, \rm{\mu s})$ is assumed. }
\label{table:parameters}
\end{table*}
%%%%%%%%%%%%%%%%%%%%%%%%%%%%%%%%%%%%%%%%%%%

In the absence of damping, exact diagonalization of the Jaynes-Cumming
Hamiltonian yields the excited eigenstates (dressed states) \cite{haroche:92}
\begin{eqnarray}
\ket{\overline{+,n}}
&=& \cos\theta_n \ket{\downarrow,n} + \sin\theta_n \ket{\uparrow,n+1}
\label{eq:dressed-states_plus}\\
\ket{\overline{-,n}} 
&=& -\sin\theta_n \ket{\downarrow,n} + \cos\theta_n \ket{\uparrow,n+1}
\label{eq:dressed-states_minus}
\end{eqnarray}
and ground state $\ket{\ua,0}$ with corresponding eigenenergies
\begin{eqnarray}
E_{\overline{\pm,n}} 
&=& (n+1)\hbar\omegar \pm \frac{\hbar}{2}\sqrt{4g^2(n+1)+\Delta^2}
\label{eq:dressed-energy_excited}\\
E_{\ua,0} &=& -\frac{\hbar\Delta}{2}.
\label{eq:dressed-energy_ground}
\end{eqnarray}
In these expressions,
\be
\theta_n = \frac{1}{2}\tan^{-1}\left(\frac{2g\sqrt{n+1}}{\Delta}\right),
\ee
and $\Delta\equiv\Omega-\omegar$ the atom-cavity detuning.

Figure \ref{fig:cQED}b) shows the spectrum of these dressed-states for the case
of zero detuning, $\Delta=0$, between the atom and the cavity.  In this situation,
degeneracy of the pair of states with $n$ quanta is lifted by $2g\sqrt{n+1}$ due to
the atom-photon interaction.  
In the manifold with a single excitation, Eqs.~\eq{eq:dressed-states_plus} and 
\eq{eq:dressed-states_minus} reduce to the maximally entangled atom-field states 
$\ket{\overline{\pm,0}} = \left(\ket{\ua,1} \pm \ket{\da,0}\right)/\sqrt{2}$.  
An initial zero-photon excited atom state $\ket{\ua,0}$ will
therefore flop into a photon $\ket{\da,1}$ and back again 
at the vacuum Rabi frequency $g/\pi$.  Since the excitation is half atom and 
half photon, the decay rate of $\ket{\overline{\pm,0}}$ is $(\kappa+\gamma)/2$.
The pair of states $\ket{\overline{\pm,0}}$ will be resolved in
a transmission experiment if the splitting $2g$
is larger than this linewidth.  The value of $g={\cal E}_{\rm rms}d/\hbar$ is
determined by the transition dipole moment $d$ and the rms
zero-point electric field of the cavity mode.  Strong coupling is
achieved when $g\gg \kappa,\gamma$ \cite{haroche:92}.

For large detuning, $g/\Delta \ll 1$, expansion of Eq.~\eq{eq:dressed-energy_excited}
yields the dispersive spectrum shown in Fig.~\ref{fig:cQED}c).  In this situation,
the eigenstates of the one excitation manifold take the form \cite{haroche:92}
\begin{eqnarray}
\ket{\overline{-,0}} &\sim& -(g/\Delta)\ket{\da,0} +\ket{\ua,1}\\
\ket{\overline{+,0}} &\sim& \ket{\da,0} + (g/\Delta) \ket{\ua,1}.
\end{eqnarray}
The corresponding decays rates are then simply given by
\begin{eqnarray}
\Gamma_{\overline{-,0}} &\simeq& (g/\Delta)^2 \gamma + \kappa\\
\Gamma_{\overline{+,0}} &\simeq& \gamma + (g/\Delta)^2 \kappa.
\label{eq:gamma_large_detuning}
\end{eqnarray}

More insight into the dispersive regime is gained by making the
unitary transformation
\begin{equation}
U=\exp{\left[\frac{g}{\Delta}(a\sigma^+ - a^\dagger\sigma^-)\right]}
\label{eq:transformation}
\end{equation}
and expanding to second order in $g$ (neglecting damping for the moment)
to obtain
\be
UHU^\dagger \approx \hbar\left[\omegar +
\frac{g^2}{\Delta}\sigma^z\right] a^\dagger a +
\frac{\hbar}{2}\left[\Omega+\frac{g^2}{\Delta}\right]\sigma^z.
\label{eq:Jaynes-Cummings-diagonal}
\ee
As is clear from this expression, the atom transition is ac-Stark/Lamb shifted 
by $(g^2/\Delta)(n+1/2)$.  Alternatively, one can interpret the ac-Stark shift as a
dispersive shift of the cavity transition by $\sigma_z
g^2/\Delta$. In other words, the atom pulls the cavity frequency by $\pm
g^2/\kappa\Delta$.

%%%%%%%%%%%%%%%%%%%%%%%%%%
% II: Our sc QED proposal
%%%%%%%%%%%%%%%%%%%%%%%%%%

\section{Circuit Implementation of Cavity QED}
\label{sec:scQED}

\begin{figure}[t]
\begin{center}
\includegraphics[%angle=90
,width=3.25in]{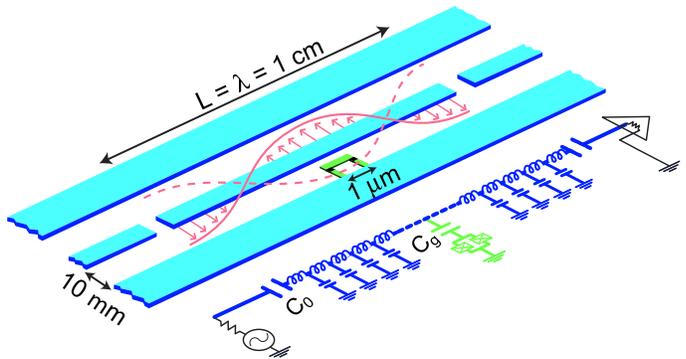} 
\caption{(color online). Schematic layout and equivalent lumped
circuit representation of proposed implementation of cavity QED using
superconducting circuits. The 1D transmission line resonator
consists of a full-wave section of superconducting coplanar
waveguide, which may be lithographically fabricated using
conventional optical lithography. A Cooper-pair box qubit is
placed between the superconducting lines, and is capacitively
coupled to the center trace at a maximum of the voltage standing
wave, yielding a strong electric dipole interaction between the
qubit and a single photon in the cavity. The box consists of two
small ($\sim 100 \, \rm{nm} \times 100 \, \rm{nm}$) Josephson
junctions, configured in a $\sim 1 \, \rm{\mu m}$ loop to permit
tuning of the effective Josephson energy by an external flux $\Phi_{\rm ext}$. 
Input and output signals are coupled to the resonator, via the
capacitive gaps in the center line, from $50 \, \Omega$
transmission lines which allow measurements of the amplitude and
phase of the cavity transmission, and the introduction of dc and
rf pulses to manipulate the qubit states. Multiple qubits (not
shown) can be similarly placed at different antinodes of the
standing wave to generate entanglement and two-bit quantum gates
across distances of several millimeters.}
\label{fig:resonatorandbox}
\end{center}
\end{figure}

We now consider the proposed realization of cavity QED using
superconducing circuits shown in Fig.~\ref{fig:resonatorandbox}.
A  1D transmission line resonator consisting of a full-wave section
of superconducting coplanar waveguide plays the role of the cavity
and a superconducting qubit plays the role of the atom.  A number of
superconducting quantum circuits could function as artificial atom,
but for definiteness we focus here on the Cooper pair box
\cite{bouchiat:98,makhlin:2001,vion:2002,lehnert:2003}.

\subsection{Cavity: coplanar stripline resonator}

An important advantage of this approach is that the zero-point
energy is distributed over a very small effective volume
($\approx 10^{-5}$ cubic wavelengths) for
our choice of a quasi-one-dimensional transmission line `cavity.'
As shown in appendix \ref{appendix:resonator}, this leads to significant 
rms voltages ${V}_{\rm rms}^0 \sim \sqrt{\hbar\omegar / c L}$ between
the center conductor and the adjacent ground plane at the anti\-nodal positions, 
where $L$ is the resonator length and $c$ is the capacitance per unit 
length of the transmission line. At a resonant frequency of $10 \, \rm{GHz}$
($h \nu / k_B \sim 0.5 \, \rm{K}$) and for a $10 \,\rm{\mu m}$ gap
between the center conductor and the adjacent ground plane,
${V}_{\rm rms} \sim 2 \, \rm{\mu V}$ corresponding to electric
fields ${\cal E}_{\rm rms} \sim 0.2 \, \rm{V/m}$, some $100$ times
larger than achieved in the 3D cavity described in
Ref.~\cite{raimond:2001}. Thus, this geometry might also be useful
for coupling to Rydberg atoms \cite{lukin:2003}.

In addition to the small effective volume, and the fact that the
on-chip realization of cQED shown in Fig.~\ref{fig:resonatorandbox}
can be fabricated with existing lithographic techniques, a
transmission-line resonator geometry offers other practical advantages
over lumped LC circuits or current-biased large
Josephson junctions. The qubit can be placed within the cavity
formed by the transmission line to strongly suppress the
spontaneous emission, in contrast to a lumped LC circuit, where
without additional special filtering, radiation and parasitic resonances
may be induced in the wiring \cite{Simmonds:2003}.
Since the resonant frequency of the transmission line is
determined primarily by a fixed geometry, its reproducibility and
immunity to 1/f noise should be superior to Josephson junction
plasma oscillators. Finally, transmission line resonances in coplanar
waveguides with $Q\sim 10^6$ have already been demonstrated
\cite{day:2003, wallraff:2003}, suggesting that the internal losses can be very
low.  The optimal choice of the resonator $Q$ in this approach is
strongly dependent on the intrinsic decay rates
of superconducting qubits which as described below, are
presently unknown, but can be determined with the setup proposed
here.  Here we assume the conservative case of
an overcoupled resonator with a $Q\sim 10^4$, which is preferable
for the first experiments.

\subsection{Artificial atom: the Cooper pair box}

\begin{figure}[b]
\begin{center}
\includegraphics[%angle=90,
width=1.75in]{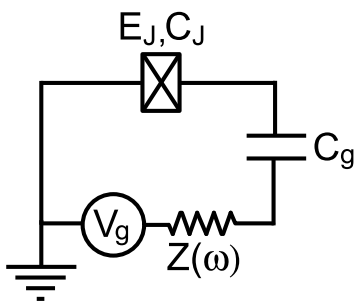} 
\caption{Circuit diagram of the Cooper pair box.  The gate voltage is connected
to the island through an environmental impedance $Z(\omega)$.}
\label{fig:qubit_out_cavity}
\end{center}
\end{figure}

Our choice of `atom', the Cooper pair box \cite{bouchiat:98,makhlin:2001}
is a mesoscopic superconducting island. As shown in Fig.~\ref{fig:qubit_out_cavity},
the island is connected to a large reservoir through a Josephson junction
with Josephson energy $E_J$ and capacitance $C_J$.  
It is voltage biased from a lead having capacitance $C_g$ to the island.
If the superconducting gap is larger than both the charging energy
$E_c = e^2/2C_\Sigma$ (where $C_\Sigma = C_J+C_g$ is the total box capacitance)
and temperature, the only relevant degree of freedom is the number
of Cooper pairs $N$ on the island.  In this basis, the Hamiltonian
describing the superconducting island takes the form
\begin{eqnarray}
H_Q &=& 4E_c \sum_N(N-N_g)^2 \ket{N}\bra{N} \nonumber\\
&& - \frac{E_J}{2} \sum_N \left( \ket{N+1}\bra{N} + h.c. \right),
\label{eq:island_hamiltonian_full}
\end{eqnarray}
where $N_g = C_gV_g/2e$
is the dimensionless gate charge representing the total polarization charge
injected into the island by the voltage source.

In the charge regime, $4E_c \gg E_J$, and restricting the gate charge
to the range $N_g \in [0,1]$, only a pair of adjacent charge states on the
island are relevant and the Hamiltonian then reduces to a $2\times 2$
matrix
\be
H_Q = -\frac{E_{\rm el}}{2} \bar\sigma^z -\frac{E_{J}}{2} \bar\sigma^x,
\label{eq:H_qubit_TLS}
\ee
with $E_{\rm el} = 4E_C (1-2N_g)$.  The Cooper pair box can in this case be
mapped to a pseudo spin-1/2 particle, with effective fields in the $x$ and
$z$ directions.

Replacing the Josephson junction by a pair of junctions in parallel each with
energy $E_J/2$, the effective field in the $x$ direction becomes
$E_J \cos(\pi\Phi_{\rm ext}/\Phi_0)/2$.
By threading a flux $\Phi_{\rm ext}$ in the loop formed by the pair of junctions and
changing the gate voltage $V_g$, it is possible to control the effective field
acting on the qubit. In the setup of Fig.~\ref{fig:resonatorandbox},
application of dc gate voltage on the island can be conveniently achieved by applying
a bias voltage to the center conductor of the transmission line.  The resonator
coupling capacitance $C_0$, the gate capacitance $C_g$
(the capacitance between the center conductor of the resonator and the island)
and the capacitance to ground of the resonator then act as a voltage divider.

\subsection{Combined system: superconducting cavity QED}

For a superconducting island fabricated inside a resonator, in addition
to a dc part $V_{\rm g}^{\rm{dc}}$, the gate voltage has a quantum part $v$.  
As shown in appendix \ref{appendix:resonator}, if the qubit is placed in
the center of the resonator, this latter contribution is given by 
$v=V^0_{\rm rms} (a^\dagger + a)$.  Taking into account both 
$V_{\rm g}^{\rm{dc}}$ and $v$ in \eq{eq:H_qubit_TLS}, we obtain
\begin{eqnarray}
H_Q
&=&
-2E_C (1-2n_g^{\rm{dc}}) 	\bar\sigma^z
- \frac{E_{J}}{2} \bar\sigma^x\nonumber\\
&&- e \frac{C_g}{C_\Sigma}\sqrt{\frac{\hbar \omegar}{Lc}}(a^\dagger+a)
(1-2N_g-\bar\sigma^z).
\label{eq:H_qubit_field_1}
\end{eqnarray}
Working in the eigenbasis $\{\ket{\ua},\ket{\da}\}$ of the first two terms 
of the above expression \cite{schoelkopf:2003},
and adding the Hamiltonian of the oscillator mode coupled to the qubit,
the Hamiltonian of the interacting qubit and resonator system takes the form
\begin{eqnarray}
\lefteqn
H & &= 
 \hbar\omegar \left(a^\dagger a + \frac{1}{2}\right)
+ \frac{\Omega}{2}\sigma^z
\label{eq:H_qubit_field_2}\\
&&- e \frac{C_g}{C_\Sigma}\sqrt{\frac{\hbar \omegar}{Lc}}
(a^\dagger+a)(1-2N_g-\cos(\theta)\sigma^z+\sin(\theta)\sigma^x).
\nonumber
\end{eqnarray}
Here, $\sigma^x$ and $\sigma^z$ are Pauli matrices in the eigenbasis
$\{\ket{\ua},\ket{\da}\}$,  $\theta = \arctan[E_J/4E_C(1-2N_g^{\rm{dc}})]$
is the mixing angle and the energy splitting of the qubit is 
$\Omega = \sqrt{E_J^2+[4E_C(1-2N_g^{\rm{dc}})]^2}$  \cite{schoelkopf:2003}.
Note that contrary to the case of a qubit fabricated outside the cavity where
the $N_g^2$ term in \eq{eq:island_hamiltonian_full} has no effect, here this
term slightly renormalize the cavity frequency $\omegar$ and displaces
the oscillator coordinate. These effects are implicit in Eq.~\eq{eq:H_qubit_field_2}.

At the charge degeneracy point (where $N_{\rm g}=C_{\rm g} V_{\rm
g}^{\rm{dc}} /2e = 1/2$ and $\theta = \pi/2$), neglecting rapidly oscillating terms
and omitting damping for the moment, Eq.~\eq{eq:H_qubit_field_2} reduces to the
Jaynes-Cummings Hamiltonian \eq{eq:Jaynes-Cummings} with $\Omega = E_J$
and the vacuum Rabi frequency
\be
g= 
\frac{\beta e}{\hbar}\sqrt{\frac{\hbar \omegar}{c L}},
\ee 
where $\beta\equiv {C_{\rm g}}/{C_\Sigma}$. The
quantum electrical circuit of Fig.~\ref{fig:resonatorandbox} is therefore
mapped to the problem of a two-level atom inside a cavity.  Away from the
degeneracy point, this mapping can still be performed, but with a coupling strength
reduced by $\sin\theta$ and an additional term proportional to
$(a^\dag+a)$.

In this circuit, the `atom' is highly polarizable at the charge degeneracy point,
having transition dipole  moment $d \equiv {\hbar g}/{{\cal E}_{\rm rms}}
\sim 2\times 10^4$ atomic units ($ea_0$), or more than an order of
magnitude larger than even a typical Rydberg atom
\cite{haroche:92}. An experimentally realistic \cite{lehnert:2003}
coupling $\beta\sim 0.1$ leads to a vacuum Rabi rate $g/\pi\sim
100$~MHz, which is three orders of magnitude larger than in
corresponding atomic microwave cQED experiments
\cite{raimond:2001}, or approximately 1\% of the transition frequency.
Unlike the usual cQED case, these artificial `atoms' remain at fixed
positions indefinitely and so do not suffer from the problem that
the coupling $g$ varies with position in the cavity. 

A comparison of the experimental parameters for implementations of
cavity QED with optical and microwave atomic systems and for the
proposed implementation with superconducting circuits is
presented in Table~\ref{table:parameters}. We assume here a relatively low
$Q=10^4$ and a worst case estimate, consistent with the bound set by previous
experiments with superconducting qubits (discussed further below),
for the intrinsic qubit lifetime of $1/\gamma \geq 2\,\rm{\mu s}$.

The standard
figures of merit~\cite{kimble:94} for strong coupling are the critical photon
number needed to saturate the atom on resonance $m_0
=\gamma^2/2g^2\le 1\times 10^{-6}$ and the minimum atom number
detectable by measurement of the cavity output $N_0
=2\gamma\kappa/g^2\le 6\times 10^{-5}$. These remarkably low
values are clearly very favorable, and show that superconducting
circuits could access the interesting regime of very strong
coupling.

%%%%%%%%%%%%%%%%%%
% III: Case of Zero Detuning
%%%%%%%%%%%%%%%%%%
\section{Zero Detuning}
\label{sec:zero_detuning}

\begin{figure}[t]
\begin{center}
\includegraphics[width=3in]{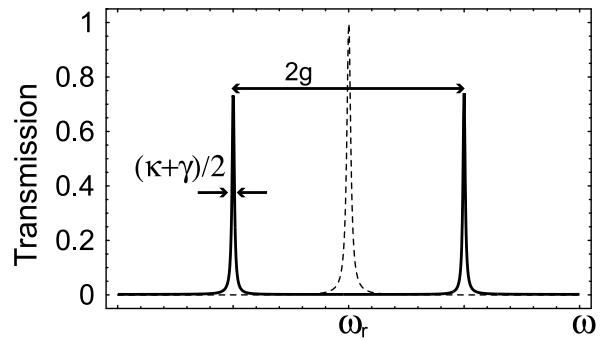} 
\caption{Expected transmission spectrum of the resonator
in the absence (broken line) and presence (full line) of a superconducting qubit
biased at its degeneracy point.  Parameters are those presented in 
Table~\ref{table:parameters}. The splitting exceeds the line width by two
orders of magnitude.}
\label{fig:resonant_spectrum}
\end{center}
\end{figure}

In the case of a low Q cavity ($g<\kappa$) and zero detuning, the radiative
decay rate of the qubit into the transmission line
becomes strongly {\em enhanced} by a factor of $Q$ relative to the
rate in the absence of the cavity \cite{haroche:92}.  This is due to the
resonant enhancement of the density of states at the atomic
transition frequency. In electrical engineering language, the
$\sim 50 \, \Omega$ external transmission line impedance is
transformed on resonance to a high value which is better matched
to extract energy from the qubit.

For strong coupling $g>\kappa,\gamma$, the first excited state becomes a doublet with
line width $(\kappa +\gamma)/2$, as explained in section
\ref{sec:review_cQED}.  As can be seen from Table~\ref{table:parameters}, the
coupling in the proposed superconducting implementation is so strong
that, even for the low $Q=10^4$ we have assumed,
$2g/(\kappa + \gamma) \sim 100$ vacuum Rabi oscillations are
possible.  Moreover, as shown in Fig.~\ref{fig:resonant_spectrum}, 
the frequency splitting ($g/\pi \sim 100 \, \rm{MHz}$) 
will be readily resolvable in the transmission spectrum of the resonator. 
This spectrum, calculated here following Ref. \cite{wang:97},
can be observed in the same manner as
employed in optical atomic experiments, with a continuous wave
measurement at low drive, and will be of practical use to find the
dc gate voltage needed to tune the box into resonance with the
cavity.

Of more fundamental importance than this simple avoided
level crossing however, is the fact that the Rabi splitting scales
with the square root of the photon number, making the level
spacing anharmonic.  This should cause a number of novel
non-linear effects \cite{walls-milburn} to appear in the spectrum
at higher drive powers when the average photon number in the
cavity is large ($\langle n \rangle > 1$).  

A conservative estimate
of the noise energy for a 10 GHz cryogenic high electron mobility
(HEMT) amplifier is $n_{\rm amp}=k_BT_N/\hbar\omega \sim 100$ photons,
where $T_N$ is the noise temperature of the amplification circuit.
As a result, these spectral features should be readily observable in a
measurement time $t_{\rm meas}= 2\ndet/\langle n
\rangle\kappa$, or only $\sim 32 \, \rm{\mu s}$ for $\langle n
\rangle\sim 1$.

%%%%%%%%%%%%%%%%%%%%%%%%%%%%%
% IV:  Case of Large Detuning I
%%%%%%%%%%%%%%%%%%%%%%%%%%%%%

\section{Large detuning: lifetime enhancement}
\label{sec:large_detuning}

For qubits {\em not} inside a cavity, fluctuation of the gate
voltage acting on the qubit is an important source of relaxation
and dephasing.  As shown in Fig.~\ref{fig:qubit_out_cavity}, in practice the 
qubit's gate is connected to the voltage source through external wiring having,
at the typical microwave transition frequency of the qubit, a 
real impedance of value close to the impedance of free space
($\sim 50 \, \Omega$).  The relaxation rate expected from purely
quantum fluctuations across this impedance (spontaneous emission) 
is~\cite{schoelkopf:2003,lehnert:2003}
\be 
\frac{1}{T_1} =
\frac{E^2_{\rm J}}{{E_{\rm J}^2+E_{\rm el}^2}}
\left(\frac{e}{\hbar}\right)^2\beta^2
S_V(+\Omega),
\label{eq:T1_vacuum_noise}
\ee
where $S_V(+\Omega) = 2\hbar\Omega \,{\rm Re}[Z(\Omega)]$
is the spectral density of voltage fluctuations across the
environmental impedance (in the quantum limit).
It is difficult in most experiments to precisely determine the real part
of the high frequency environmental impedance presented
by the leads connected to the qubit, but reasonable estimates
\cite{lehnert:2003} yield values of $T_1$ in the range of $1 \,
\rm{\mu s}$.

For qubits fabricated inside a cavity, the noise across the
environmental impedance does not couple directly to the qubit,
but only indirectly through the cavity.    
For the case of strong detuning, coupling of the qubit
to the continuum is therefore substantially reduced.  
One can view the effect of the detuned
resonator as filtering out the vacuum noise at the qubit
transition frequency or, in electrical engineering terms, as
providing an impedance transformation which strongly {\em reduces}
the real part of the environmental impedance seen by the qubit.

Solving for the normal modes of the resonator and transmission
lines, including an input impedance $R$ at each end of the resonator,
the spectrum of voltage fluctuations as seen by the qubit
fabricated in the center of the resonator can be shown to be well
approximated by
\be
S_V(\Omega) = \frac{2\hbar\omegar}{Lc}
\frac{\kappa/2}{\Delta^2+(\kappa/2)^2}.
\ee
Using this transformed
spectral density in \eq{eq:T1_vacuum_noise} and assuming a large detuning
between the cavity and the qubit, the relaxation rate due to vacuum
fluctuations takes a form that reduces to
$1/T_1\equiv \gamma_\kappa = (g/\Delta)^2\kappa \sim 1/(64 \, \rm{\mu s})$,
at the qubit's degeneracy point. This is the result already obtained in
Eq.~\eq{eq:gamma_large_detuning} using the dressed state
picture for the coupled atom and cavity, except for the additional
factor  $\gamma$ reflecting loss of energy to modes outside
of the cavity.  For large detuning, damping
due to spontaneous emission can be much less than $\kappa$.

One of the important motivations for this cQED experiment is to
determine the various contributions to the qubit decay rate
so that we can understand their fundamental physical
origins as well as engineer improvements. Besides $\gamma_\kappa$
evaluated above, there are two additional contributions to the total
damping rate
$\gamma=\gamma_\kappa+\gamma_\perp+\gamma_{\rm NR}$. Here
$\gamma_\perp$ is the decay rate into photon modes other than the
cavity mode, and $\gamma_{\rm NR}$ is the rate of other (possibly
non-radiative) decays. Optical cavities are relatively open and
$\gamma_\perp$ is significant, but for 1D microwave cavities,
$\gamma_\perp$ is expected to be negligible (despite the very
large transition dipole). For Rydberg atoms the two qubit states
are both highly excited levels and $\gamma_{\rm NR}$ represents
(radiative) decay out of the two-level subspace. For Cooper pair
boxes, $\gamma_{\rm NR}$ is completely unknown at the present
time, but could have contributions from phonons, two-level systems
in insulating \cite{Simmonds:2003} barriers and substrates, or
thermally excited quasiparticles.

For Cooper box qubits {\em not} inside a cavity, recent
experiments \cite{lehnert:2003} have determined a relaxation time
$1/\gamma=T_{1}\sim 1.3 \, \rm{\mu s}$ despite the back action of
continuous measurement by a SET electrometer.  Vion et al.
\cite{vion:2002} found $T_{1}\sim 1.84 \, \rm{\mu s}$ (without
measurement back action) for their charge-phase qubit.  
Thus in these experiments, if there are non-radiative
decay channels, they are at most comparable to the vacuum
radiative decay rate (and may well be much less) estimated using
Eq.~\eq{eq:T1_vacuum_noise}. Experiments with a cavity will present
the qubit with a simple and well controlled
electromagnetic environment, in which the radiative lifetime can
be enhanced with detuning to $1/\gamma_\kappa > 64 \, \rm{\mu s}$,
allowing $\gamma_{\rm NR}$ to dominate and yielding valuable
information about any non-radiative processes.

%%%%%%%%%%%%%%%%%%%%%%%%%%%%%
%  V: Case of Large Detuning II
%%%%%%%%%%%%%%%%%%%%%%%%%%%%%
\section{Dispersive QND Readout of Qubit}
\label{sec:read-out}

In addition to lifetime enhancement, the dispersive regime is
advantageous for read-out of the qubit.  This can be realized by
microwave irradiation of the cavity and then probing the transmitted
or reflected photons~\cite{note_jena}.

\subsection{Measurement Protocol}

A drive of frequency $\omega_{\mu \rm w}$ on the resonator can be modeled
by \cite{haroche:92}
\be
H_{\mu w}(t) = 
\hbar\varepsilon(t) (a^\dag e^{-i\omegad} + a e^{+i\omegad}),
\label{eq:Hamiltonian_drive}
\ee
where $\varepsilon(t)$ is a measure of the drive the amplitude.  In the dispersive limit, 
one expects from Fig.~\ref{fig:cQED}c) peaks in the transmission spectrum
at $\omegar-g^2/\Delta$ and $\Omega + 2g^2/\Delta$ if the qubit is initially
in its ground state. In a frame rotating at the drive frequency, the matrix elements
for these transitions are respectively
\begin{eqnarray}
\bra{\ua,0}  H_{\mu w} \ket{\overline{-,n}}  &\sim& \varepsilon
\nonumber\\
\bra{ \ua,0 } H_{\mu w} \ket{\overline{+,n}} &\sim& \frac{\varepsilon g}{\Delta}.
\label{eq:matrix_element_drive}
\end{eqnarray}
In the large detuning case, the peak at $\Omega + 2g^2/\Delta$,
corresponding approximatively to a qubit flip, is highly suppressed. 

\begin{figure}[t]
\begin{center}
\includegraphics[%angle=90,
width=3in]{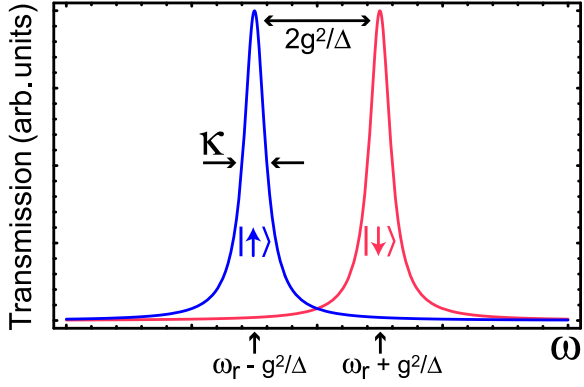}
\caption{(color online). Transmission spectrum of
the cavity, which is ``pulled" by an amount $\pm g^2/\Delta = 2.5
\times 10^{-4}\times \omegar$, depending on the state of the
qubit (red for the excited state, blue for the ground state).  To
perform a measurement of the qubit, a pulse of microwave photons,
at a probe frequency $\omegad = \omegar$ or $\omegar\pm g^2/\Delta$
is sent through the cavity.
Additional peaks near $\Omega$ corresponding to qubit flips are suppressed by
$g/\Delta$.}
\label{fig:spectrum}
\end{center}
\end{figure}

The matrix element corresponding to a qubit flip from the excited
state is also suppressed and, as shown in Fig.~\ref{fig:spectrum},
depending on the qubit being in its ground or excited states, the
transmission spectrum will present a peak of width $\kappa$
at $\omegar-g^2/\Delta$  or $\omegar+g^2/\Delta$.
With the parameters of Table~\ref{table:parameters}, this dispersive pull
of the cavity frequency is $\pm g^2/\kappa\Delta=\pm 2.5$ line widths for
a 10\% detuning.   Exact diagonalization \eq{eq:dressed-energy_excited} shows
that the pull is power dependent and decreases in magnitude for cavity photon
numbers on the scale $n=n_{\rm crit}\equiv \Delta^2/4g^2$.
In the regime of non-linear response, single-atom optical
bistability \cite{walls-milburn} can be expected when the drive
frequency is off resonance at low power but on resonance at high
power \cite{smgunpublished}.

The state-dependent pull of the cavity frequency by the qubit can
be used to entangle the state of the qubit with that of the
photons transmitted or reflected by the resonator.  For $g^2/\kappa\Delta > 1$,
as in Fig.~\ref{fig:spectrum}, the pull is greater than the line width and
irradiating the cavity at one of the pulled frequencies
$\omegar \pm g^2/\Delta$, the transmission of the cavity will be close
to unity for one state of the qubit and close to zero for the other \cite{note_pull}.

Choosing the drive to be instead at the bare cavity frequency
$\omegar$, the state of the qubit is encoded in the phase of 
the reflected and transmitted microwaves. An initial qubit state 
$\ket{\chi} = \alpha\ket{\uparrow}+\beta
\ket{\downarrow}$ evolves under microwave
irradiation into the entangled state
$\ket{\psi}=\alpha \ket{\uparrow,\theta}
+ \beta\ket{\downarrow,-\theta}$, where $\tan\theta =
{2g^2}/{\kappa\Delta}$, and $\ket{\pm\theta}$ are
(interaction representation) coherent states with the appropriate
mean photon number and opposite phases.  In the situation
where ${g^2}/{\kappa\Delta} \ll 1$, this is the most appropriate
strategy.

It is interesting to note that such an entangled state
can be used to couple qubits in distant resonators and allow
quantum communication \cite{vanEnk:98}. Moreover, if an independent
measurement of the qubit state can be made, such states can
be turned into photon Schr\"odinger cats \cite{haroche:92}.

To characterize these two measurement schemes corresponding
to two different choices of the drive frequency,
we compute the average photon number inside the resonator $\bar n$
and the homodyne voltage on the 50$\Omega$ impedance
at the output of the resonator.  Since the power coupled to the
outside of the resonator is 
$P = \langle n\rangle\hbar\omegar\kappa/2 = \langle V_{\rm out}\rangle^2/R$,
the homodyne voltage can be expressed as
$\langle V_{\rm out}\rangle = \sqrt{R\hbar\omegar\kappa} \langle a+a^\dag\rangle/2$ 
and is proportional to the real part of the field inside the cavity.

In the absence of dissipation, the time dependence of the field inside the
cavity can be obtained in the Heisenberg picture from 
Eqs.~\eq{eq:Jaynes-Cummings-diagonal} and \eq{eq:Hamiltonian_drive}.
This leads to a closed set of differential equations for $a$, $\sigma_z$ and
$a\sigma_z$ which is easily solved.  In the presence of dissipation however (i.e.
performing the transformation~\eq{eq:transformation} on $H_{\kappa}$
and $H_{\gamma}$, and adding the resulting terms to 
Eqs.~\eq{eq:Jaynes-Cummings-diagonal} and \eq{eq:Hamiltonian_drive}),
the set is no longer closed and we resort to numerical 
stochastic wave function calculations \cite{schack:97}.

\begin{figure}[t]
\begin{center}
\includegraphics[width=2.5in]{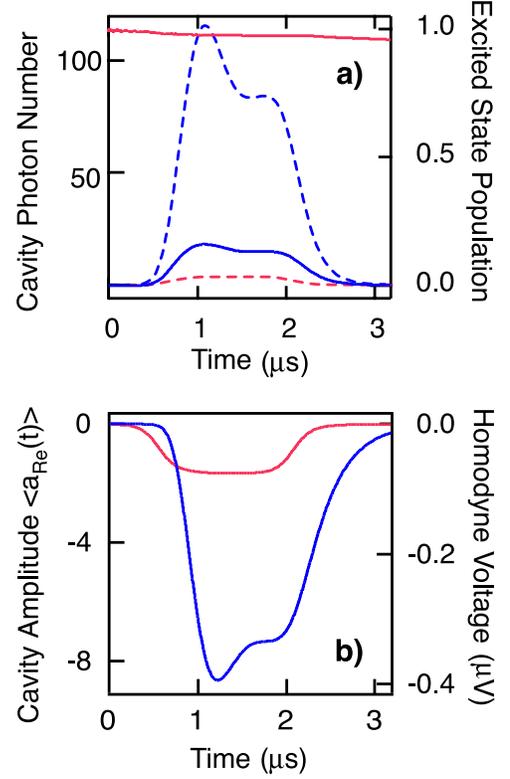}
\caption{(color online). Results of numerical simulations using
the quantum state diffusion method. A microwave pulse of duration 
$\sim 15/\kappa$ and centered at the pulled frequency
$\omegar + g^2/\Delta$ drives the cavity.
a) The occupation probability of the excited state (right axis), for the
case in which the qubit is initially in the ground (blue) or
excited (red) state and intracavity photon number (left axis),
are shown as a function of time. Though the
qubit states are temporarily coherently mixed during the pulse, the
probability of real transitions is seen to be small. Depending on
the qubit's state, the pulse is either on or away from the combined
cavity-qubit resonance, and therefore is mostly transmitted or mostly reflected.
b) The real component of the cavity electric field amplitude (left
axis), and the transmitted voltage phasor (right axis) in the
output transmission line, for the two possible initial qubit states.
The parameters used for the simulation are presented in
Table~\ref{table:parameters}.}
\label{fig:stochastic_number}
\end{center}
\end{figure}

\begin{figure}[t]
\begin{center}
\includegraphics[width=2.5in]{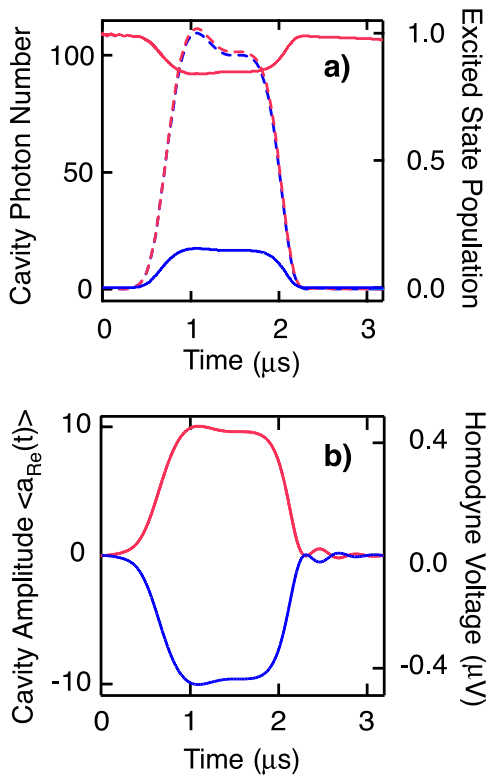}
\caption{(color online). Same as Fig.~\ref{fig:stochastic_number}
for the drive at the bare cavity frequency $\omegar$.
Depending on the qubit's state, the pulse is either above or below 
the combined cavity-qubit resonance, and so is partly transmitted
and reflected but with a large relative phase shift that can be
detected with homodyne detection. 
In b), the opposing phase shifts cause a change in sign of the output,
which can be measured with high signal-to-noise to realize a
single-shot, QND measurement of the qubit.}
\label{fig:stochastic_phase}
\end{center}
\end{figure}

Figures~\ref{fig:stochastic_number} and \ref{fig:stochastic_phase} show
the numerical results for the two choices of drive frequency and using
the parameters of Table~\ref{table:parameters}.
For these calculations, a pulse of duration $\sim15/\kappa$
with a hyperbolic tangent rise and fall, is used to excite the cavity.  
Fig.~\ref{fig:stochastic_number} corresponds to a drive at the
pulled frequency $\omegar+g^2/\Delta$. In Fig.~\ref{fig:stochastic_number}a)
the probability $P_\da$ to find the qubit in its excited state (right axis) is plotted
as a function of time for the qubit initially in the ground (blue) or excited state (red). 
The dashed lines represent the corresponding number of photons in the cavity 
(left axis).
Fig.~\ref{fig:stochastic_number}b) shows, in a frame rotating at the drive frequency,
the real part of the cavity electric field amplitude (left axis) and transmitted
voltage phase (right axis) in the output transmission line,
again for the two possible initial qubit states.  These quantities are shown
in Fig.~\ref{fig:stochastic_phase} for a drive at the bare frequency $\omegar$.

As expected, for the first choice of drive frequency, the information about the state
of the qubit is mostly stored in the number of transmitted photons.  When the drive
is at the bare frequency however, there is very little information in the photon
number, with most of the information being stored in the phase of the transmitted and
reflected signal. This phase shift can be measured using standard heterodyne
techniques.  
As also discussed in appendix~\ref{appendix:QND},
both approaches can serve  as a high efficiency quantum 
non-demolition dispersive readout of the state of the qubit.

\subsection{Measurement Time and Backaction}

As seen from Eq.~\eq{eq:Jaynes-Cummings-diagonal}, the 
back action of the dispersive cQED measurement is due to quantum
fluctuations of the number of photons $n$ within the cavity.  These fluctuations
cause variations in the ac Stark shift $(g^2/\Delta)n\sigma^z$ that
in turn dephase the qubit.   It is useful to compute the corresponding
dephasing rate and compare it with the measurement
rate, i.e. the rate at which information about the state of the qubit
can be acquired.

To determine the dephasing rate, we assume that the cavity is driven
at the bare cavity resonance frequency and that the pull of the resonance
is small compared to the line width $\kappa$.
The relative phase accumulated between the ground and excited states
of the qubit is
\be
\varphi(t) = 2 \frac{g^2}{\Delta} \int_0^t dt' n(t')
\ee
which yield a mean phase advance $\langle\varphi\rangle = 2\theta_0 N$
with $\theta_0 = 2g^2/\kappa\Delta$ and  $N= \kappa\bar n t/2$
the total number of transmitted photons \cite{walls-milburn}.  For weak
coupling, the dephasing time will greatly exceed $1/\kappa$ and, in the
long time limit, the noise in $\varphi$ induced by the ac Stark shift will
be gaussian.  Dephasing can then be evaluated by computing the
long time decay of the correlator
\begin{eqnarray}
\langle \sigma^+(t)\sigma^-(0)\rangle
&=& \langle e^{i\int_0^t dt'\varphi(t')} \rangle \nonumber\\
&\simeq& e^{-\frac{1}{2} \left(2\frac{g^2}{\Delta}\right)^2 
\int_0^t\int_0^t dt_1dt_2 \langle n(t_1) n(t_2)\rangle}.
\label{eq:dephasing_correlator}
\end{eqnarray}
To evaluate this correlator in the presence of a continuous-wave (CW) drive
on the cavity, we first perform a canonical transformation on the cavity
operators $a^{(\dag)}$ by writing them in terms of a classical
$\alpha^{(*)}$ and a quantum part $d^{(\dag)}$:
\be
a(t)= \alpha(t) + d(t).
\ee
Under this transformation, the coherent state obeying
$a\ket{\alpha} = \alpha\ket{\alpha}$, is simply the vacuum
for the operator $d$.  It is then easy to verify that
\be
\langle (n(t)-\bar n)(n(0)-\bar n)\rangle
= \alpha^2 \langle d(t)d^\dag(0)\rangle
= \bar n e^{-\frac{\kappa}{2}\mid t \mid}.
\ee
It is interesting to note that the factor of $1/2$ in the exponent is due to the
presence of the coherent drive.  If the resonator is not driven, the photon
number correlator rather decays at a rate $\kappa$.  Using this
result in \eq{eq:dephasing_correlator} yields the dephasing rate
\be
%\frac{1}{T_\varphi}
\Gamma_\varphi =  4 \theta_0^2\frac{\kappa}{2}\bar n.
\label{eq:dephasing_rate}
\ee
Since the rate of transmission on resonance is $\kappa\bar n/2$, this
means that the dephasing per transmitted photon is $4\theta_0^2$.

To compare this result to the measurement time $T_{\rm meas}$, we
imagine a homodyne measurement to determine the transmitted phase.
Standard analysis of such an interferometric set up
\cite{walls-milburn} shows that the minimum phase change which can
be resolved using $N$ photons is $\delta\theta=1/\sqrt{N}$.  Hence
the measurement time to resolve the phase change $\delta\theta =
2\theta_0$ is
\be
T_{\rm m} = \frac{1}{2\kappa{\bar n}\theta_0^2},
\ee
which yields
\be
T_{\rm m}\Gamma_\varphi =1.
\ee
This exceeds the quantum limit \cite{devoret:2000} $T_{\rm
m}\Gamma_\varphi =1/2$ by a factor of 2.  Equivalently, in the
language of Ref.~\cite{clerk:2003} (which uses a definition of the
measurement time twice as large as that above) the efficiency
ratio is $\chi \equiv 1/(T_{\rm m}\Gamma_\varphi) = 0.5$.

The failure to reach the quantum limit can be traced
\cite{marquardt:2003} to the fact that that the coupling of the
photons to the qubit is not adiabatic.  A small fraction
$R\approx\theta_0^2$ of the photons incident on the resonator are
reflected rather than transmitted.  Because the phase shift of the
reflected wave \cite{walls-milburn} differs by $\pi$ between the
two states of the qubit, it turns out that, despite its weak
intensity, the reflected wave contains precisely the same amount
of information about the state of the qubit as the transmitted
wave which is more intense but has a smaller phase shift.  In the
language of Ref.~\cite{clerk:2003}, this `wasted' information
accounts for the excess dephasing relative to the measurement
rate.  By measuring also the phase shift of the reflected photons,
it could be possible to reach the quantum limit. 

Another form of possible back action is mixing transitions 
between the two qubit states induced by the microwaves.  First,
as seen from Fig.~\ref{fig:stochastic_number}a)
and \ref{fig:stochastic_phase}a), increasing the average number of
photons in the cavity induces mixing.
This is simply caused by dressing of the qubit by the cavity photons. 
Using the dressed states \eq{eq:dressed-states_plus} and
\eq{eq:dressed-states_minus}, the level of this coherent mixing 
can be estimated as
\begin{eqnarray}
P_{\da,\ua}
&=& \frac{1}{2}\bra{\overline{\pm,n}}\id\pm\sigma^z\ket{\overline{\pm,n}}\\
&=& \frac{1}{2}\left(1\pm \frac{\Delta}{\sqrt{4g^2(n+1)+\Delta^2}}\right)
\end{eqnarray}
Exciting the cavity to $n=\ncrit$, yields $P_\da \sim 0.85$.  As is clear from the
numerical results, this process is completely reversible and does not lead
to errors in the read-out.

The drive can also lead to real transitions between the qubit states.
 However, since the coupling is so strong, large detuning
$\Delta = 0.1 \, \omegar$ can be chosen,
making the mixing rate limited not by the frequency spread of the
drive pulse, but rather by the width of the qubit excited state
itself. The rate of driving the qubit from ground to excited state
when $n$ photons are in the cavity is $R\approx
n(g/\Delta)^2\gamma$. If the measurement pulse excites the cavity
to $n=\ncrit$, we see that the excitation rate is still only 1/4
of the relaxation rate.  As a result, the main limitation on the fidelity of
this QND readout is the decay of the excited state of the qubit
during the course of the readout. This occurs (for small $\gamma$)
with probability $P_{\rm relax}\sim \gamma t_{\rm meas} \sim 15
\times \gamma/\kappa \sim 3.75 \, \%$ and leads to a small error
$P_{\rm err} \sim 5 \gamma/\kappa \sim 1.5 \, \%$ in the measurement,
 where we have taken $\gamma = \gamma_\kappa$. As confirmed by the numerical
calculations of Fig.~\ref{fig:stochastic_number} and \ref{fig:stochastic_phase}, 
this dispersive measurement is therefore highly non-demolition.

\subsection{Signal-to-Noise}

%%%%%%%%%%%%%%%%%%%%%%%%%%%%%%%%%%%%%%%%%%%
\begin{table*}[ht]
\begin{center}
\begin{small}
\begin{tabular}{|l|c|c|}\hline
parameter&symbol&1D circuit\\ \hline %
dimensionless cavity pull&$g^2/\kappa\Delta$&2.5\\ \hline %
cavity-enhanced lifetime&$\gamma^{-1}_\kappa=(\Delta/g)^2\kappa^{-1}$&$64\ \mu$s \\ \hline
%%readout excitation&$P_{{\rm mix}\uparrow}=n_{\rm det}(g/\Delta)^2\gamma/\kappa$&$0.02; 6 \times 10^{-4}$\\ \hline %
%readout de-excitaton&$P_{{\rm mix}\downarrow}=\gamma/\kappa$&$0.08;\, 2.5\times 10^{-3}$\\ \hline %
readout SNR&SNR = $(\ncrit/\ndet)\kappa/2\gamma$ &200 (6) \\ \hline %
readout error&$P_{\rm err} \sim 5 \times \gamma/\kappa$ & 1.5 \% (14\%)\\ \hline %
1 bit operation time& $T_\pi> 1/\Delta$ & $> 0.16 \, \rm{ns}$\\ \hline %
entanglement time&$t_{\sqrt{i{\rm SWAP}}} =\pi\Delta/4g^2$&$\sim 0.05\ \mu$s\\ \hline %
2 bit operations&$N_{\rm op}= 1/[\gamma\, t_{\sqrt{i{\rm SWAP}}}]$&$> 1200$ (40)\\ \hline %
\end{tabular}
\end{small}
\end{center}
\caption{Figures of merit for readout and multi-qubit entanglement
of superconducting qubits using dispersive (off-resonant) coupling
to a 1D transmission line resonator. The same parameters as Table
1, and a detuning of the Cooper pair box from the resonator of
10\% ($\Delta = 0.1 \, \omegar$), are assumed. Quantities
involving the qubit decay $\gamma$ are computed both for the
theoretical lower bound $\gamma = \gamma_\kappa$ for spontaneous
emission via the cavity, and (in parentheses) for the current
experimental upper bound $1/\gamma \ge 2 \, \rm{\mu s}$. Though
the signal-to-noise of the readout is very high in either case,
the estimate of the readout error rate is dominated by the
probability of qubit relaxation during the measurement, which has
a duration of a few cavity lifetimes ($\sim 1-10 \ \kappa^{-1}$).
If the qubit non-radiative decay is low, both high efficiency
readout and more than $10^3$ two-bit operations could be
attained.}
\label{table:fig_merit}
\end{table*}
%%%%%%%%%%%%%%%%%%%%%%%%%%%%%%%%%%%%%%%%%%%

For homodyne detection in the case where the cavity pull $g^2/\Delta\kappa$
is larger than one, the signal-to-noise ratio (SNR) is given by the ratio of the
number of photons $n_{\rm sig} = n \kappa \Delta t   /2$ accumulated
over an integration period $\Delta t$, divided by the detector noise
$n_{\rm amp} = k_BT_N/\hbar\omegar$.  Assuming the integration
time to be limited by the qubit's decay time $1/\gamma$ and
exciting the cavity to a maximal amplitude $\ncrit=100\sim \ndet$,
we obtain SNR = $(\ncrit/\ndet)(\kappa/2\gamma)$.  If the qubit lifetime
is longer than a few cavity decay times ($1/\kappa = 160 \,\rm{ns}$),
this SNR can be very large.  In the most optimistic situation where
$\gamma = \gamma_\kappa$, the signal-to-noise ratio
is SNR=200.

When taking into account the fact that the qubit has a finite probability to
decay during the measurement, a better
strategy than integrating the signal for a long time is to take advantage
of the large SNR to measure quickly.
Simulations have shown that in the situation where 
$\gamma = \gamma_\kappa$, the optimum integration
time is  roughly 15 cavity lifetimes.  This is the pulse length used
for the stochastic numerical simulations shown above.
The readout fidelity, including the effects of
this stochastic decay, and related figures of merit of the single-shot
high efficiency QND readout are summarized in Table~\ref{table:fig_merit}. 

This scheme has other interesting features that are worth mentioning
here.  First, since nearly all the energy used in this dispersive measurement
scheme is dissipated in the remote terminations of the input and output
transmission lines, it has the practical advantage of avoiding quasiparticle
generation in the qubit.

Another key feature of the cavity QED readout is that it lends
itself naturally to operation of the box at the charge degeneracy
point ($N_g=1/2$), where it has been shown that $T_2$ can be
enormously enhanced \cite{vion:2002} because the energy splitting
has an extremum with respect to gate voltage and isolation of the
qubit from 1/f dephasing is optimal. The derivative of the energy
splitting with respect to gate voltage is the charge difference in
the two qubit states.  At the degeneracy point this derivative
vanishes and the environment cannot distinguish the two states and
thus cannot dephase the qubit. This also implies that a charge
measurement cannot be used to determine the state of the system
\cite{armour:2002,elinor:2003}. While the first derivative of the
energy splitting with respect to gate voltage vanishes at the
degeneracy point, the second derivative, corresponding to the
difference in charge {\em polarizability} of the two quantum
states, is {\em maximal}.  One can think of the qubit as a
non-linear quantum system having a state-dependent capacitance (or
in general, an admittance) which changes sign between the ground
and excited states \cite{averin:2003}. It is this change in polarizability which 
is measured in the dispersive QND measurement. 

In contrast, standard charge measurement schemes \cite{nakamura:99,lehnert:2003}
require moving away from the optimal point.
Simmonds et al. \cite{Simmonds:2003} have recently
raised the possibility that there are numerous parasitic
environmental resonances which can relax the qubit when its
frequency $\Omega$ is changed during the course of moving the
operating point.  The dispersive cQED measurement is therefore
highly advantageous since it operates best at the charge
degeneracy point.  In general, such a measurement of an ac
property of the qubit is strongly desirable in the usual case
where dephasing is dominated by low frequency (1/f) noise. Notice
also that the proposed quantum non-demolition measurement would be
the inverse of the atomic microwave cQED measurement in which the
state of the photon field is inferred non-destructively from the
phase shift in the state of atoms sent through the cavity
\cite{raimond:2001}.

\section{Coherent control}
\label{sec:one_qubit}

\begin{figure}[b]
\begin{center}
\includegraphics[width=2.5in]{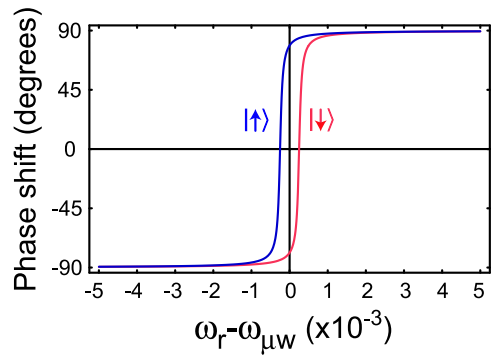} 
\caption{
(color online).
Phase shift of the cavity field for the two states of the qubit as a function
of detuning between the driving and resonator frequencies.
Obtained from the steady-state solution of the equation of motion
for $a(t)$ while only taking into account damping on the cavity  and
using the parameters of Table~\ref{table:parameters}.
Read-out of the qubit is realized at, or close to, zero detuning between
the drive and resonator frequencies where the dependence of the phase
shift on the qubit state is largest.  
Coherent manipulations of the qubit are realized close to the qubit
frequency which is 10\% detuned from the cavity (not shown on this scale).
At such large detunings, these is little dependence of the phase shift
on the qubit's state.
}
\label{fig:phase_shift}
\end{center}
\end{figure}

While microwave irradiation of the cavity at its resonance frequency
constitutes a measurement, irradiation close to the qubit's
frequency can be used to coherently control the state of the qubit.
In the former case, the phase shift of the transmitted wave is
strongly dependent on the state of the qubit and hence the photons
become entangled with the qubit,  as shown in Fig.~\ref{fig:phase_shift}. 
In the latter case however, driving is {\em not} a measurement because, 
for large detuning, the photons are largely reflected with a phase shift
which is independent of the state of the qubit.  There is therefore
little entanglement between the field and the qubit in this situation
and the rotation fidelity is high.

To model the effect of the drive on the qubit, we add the microwave drive of
Eq.~\eq{eq:Hamiltonian_drive}
to the Jaynes-Cumming Hamiltonian \eq{eq:Jaynes-Cummings} and
apply the transformation \eq{eq:transformation} (again neglecting damping) 
to obtain the following effective one-qubit Hamiltonian
\begin{eqnarray}
H_{1q} & = & 
\frac{\hbar}{2}
\left[\Omega + 2\frac{g^2}{\Delta}\left(a^\dag a +\frac{1}{2}\right) - \omega_{\mu w}\right]
\sigma^z +
\hbar\frac{g\varepsilon(t)}{\Delta}\sigma^x
\nonumber\\
&&+\hbar(\omegar - \omega_{\mu w}) a^\dag a 
+  \hbar\varepsilon(t) (a^\dag+ a),
\end{eqnarray}
in a frame rotating at the drive frequency $\omega_{\mu w}$.  Choosing
$\omega_{\mu w} =  \Omega + (2n+1)g^2/\Delta$, $H_{1q}$ generates rotations
of the qubit about the $x$ axis with Rabi frequency $g\varepsilon/\Delta$.
Different drive frequencies can be chosen to realize rotations around
arbitrary axes in the $x$--$z$ plane.  In particular, choosing
$\omega_{\mu w} =  \Omega + (2n+1)g^2/\Delta -  2g\varepsilon/\Delta$
and $t=\pi\Delta/2\sqrt{2}g\varepsilon$ generates the Hadamard transformation
$\mathcal{H}$.  Since $\mathcal{H}\sigma_x\mathcal{H}=\sigma_z$,
these two choices of frequency are sufficient to realize any 1-qubit
logical operation.

Assuming that we can take full advantage of lifetime enhancement
inside the cavity (i.e.~that $\gamma = \gamma_\kappa$),
the number of $\pi$ rotations about the $x$ axis which can be carried out is 
$N_{\rm \pi} = 2\varepsilon\Delta/\pi g\kappa\sim 10^5\varepsilon$
for the experimental parameters assumed in Table~\ref{table:parameters}.
For large $\varepsilon$, the choice of drive frequency must take into account
the power dependence of the cavity frequency pulling.

\begin{figure}[t]
\begin{center}
\includegraphics[width=2.5in]{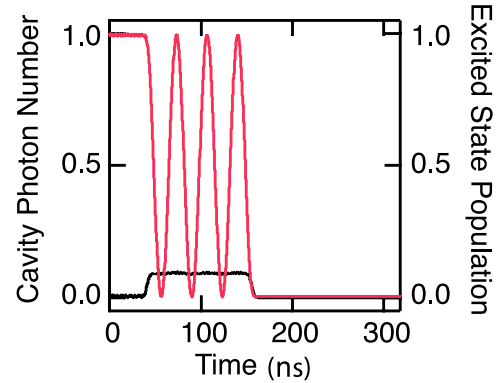} 
\caption{(color online).
Numerical stochastic wave function simulation showing coherent control of
a qubit by microwave irradiation of the cavity at the ac-Stark and Lamb shifted
qubit frequency.  The qubit is first left to evolve freely for
about 40ns.  The drive is turned on for $t=7\pi\Delta/2g\varepsilon\sim$115ns,
corresponding to 7$\pi$ pulses, and then turned off.  Since the drive
is tuned far away from the cavity, the cavity photon number is small even for
the moderately large drive amplitude $\varepsilon = 0.03\omegar$ used here.
}
\label{fig:bit-flip}
\end{center}
\end{figure}

Numerical simulation shown in Fig.~\ref{fig:bit-flip} confirms this
simple picture and that single-bit rotations can be performed with very high fidelity.
It is interesting to note that since detuning between the resonator and the drive
is large, the cavity is only virtually populated, with an average photon number
$\bar n \approx \varepsilon^2/\Delta^2 \sim 0.1$.  
Virtual population and depopulation of the cavity can be realized much faster
than the cavity lifetime $1/\kappa$ and, as a result, the qubit feels the effect of
the drive rapidly after the drive has been turned on. The limit on the speed
of turn on and off of the drive is set by the detuning $\Delta$.  If the drive
is turned on faster than $1/\Delta$, the frequency spread of the drive is
such that part of the drive's photons will pick up phase information
(see Fig.~\ref{fig:phase_shift}) and dephase the qubit.  
As a result, for large detuning, this approach leads to a 
fast and accurate way to coherently control the state of the qubit.

To model the effect  of the drive on the resonator an alternative model is to
use the cavity-modified Maxwell-Bloch equations \cite{wang:97}.
As expected, numerical integration of the Maxwell-Bloch equations reproduce
very well the stochastic numerical results when the drive is at the
qubit's frequency but do {\em not} reproduce these numerical results 
when the drive is close to the bare resonator frequency
(Fig.~\ref{fig:stochastic_number} and \ref{fig:stochastic_phase}),
i.e. when entanglement between the qubit and the photons cannot be neglected.

%%%%%%%%%%%%%%%%%%%%%%%%%%%%%%%%%%%%%%%%%%%%%%%%%%%%%%%%%%%%%%%%%
%  VI:  Multiple qubits
%%%%%%%%%%%%%%%%%%%%%%%%%%%%%%%%%%%%%%%%%%%%%%%%%%%%%%%%%%%%%%%%%

\section{Resonator as Quantum Bus: Entanglement of Multiple Qubits}
\label{sec:many_qubits}

The transmission-line resonator has the advantage that it
should be possible to place multiple qubits along its length
($\sim 1 \, \rm{cm}$) and entangle them together, which is an
essential requirement for quantum computation.  For the case of two
qubits, they can be placed closer to the ends of the resonator but
still well isolated from the environment and can be separately dc
biased by capacitive coupling to the left and right center
conductors of the transmission line. Additional qubits would
have to have separate gate bias lines installed.

For the pair of qubits labeled $i$ and $j$, both coupled with strength $g$ 
to the cavity and detuned from the resonator but in resonance with 
each other, the transformation \eq{eq:transformation} yields the effective  
two-qubit Hamiltonian \cite{raimond:2001,sorensen:99,zheng:2000}
\begin{eqnarray}
H_{2q} & \approx 
& \hbar\left[\omegar + \frac{g^2}{\Delta}(\sigma_i^z+\sigma_j^z) \right] a^\dagger a
\\
& &  + \frac{1}{2}\hbar\left[\Omega+\frac{g^2}{\Delta}\right](\sigma_i^z+\sigma_j^z)
+ \hbar\frac{g^2}{\Delta}(\sigma_i^+\sigma_j^-  + \sigma_i^-\sigma_j^+).
\nonumber
\label{eq:2qubits}
\end{eqnarray}
In addition to ac-Stark and Lamb shifts, the last term couples the
qubits thought virtual excitations of the resonator.

In a frame rotating at the qubit's frequency $\Omega$, $H_{2q}$ generates
the evolution
\begin{eqnarray}
U_{2q}(t) &=&
\exp
\left[
-i\frac{g^2}{\Delta}t
\left(a^\dag a +\frac{1}{2}\right)
\left(\sigma_i^z+\sigma_j^z\right)
\right]
\nonumber\\
&&\quad\cdot\begin{pmatrix}
1&&&\\
&\cos{\frac{g^2}{\Delta}t}&i\sin{\frac{g^2}{\Delta}t}&\\
&i\sin{\frac{g^2}{\Delta}t}&\cos{\frac{g^2}{\Delta}t}&\\
&&&1
\end{pmatrix}\otimes\id_r,
\end{eqnarray}
where $\id_r$, is the identity operator in the resonator space.
Up to phase factors, this corresponds at $t =\pi\Delta/4g^2\sim 50 \, \rm{ns}$
to a $\sqrt{i\rm SWAP}$ logical operation.  Up to one-qubit gates, this operation is
equivalent to the controlled-NOT.  Together with one-qubit gates, the interaction
$H_{2q}$ is therefore sufficient for universal quantum computation \cite{barenco:95}.
Assuming again that we can take full advantage of
the lifetime enhancement inside the cavity, the number of $\sqrt{i\rm SWAP}$
operations which can be carried out is $N_{\rm 2q} = 4\Delta/\pi\kappa\sim 1200$
for the parameters assumed above. This can be further
improved if the qubit's non-radiative decay is sufficiently small, and
higher $Q$ cavities are employed.

When the qubits are detuned from each other, the off-diagonal coupling
provided by $H_{2q}$ is only weakly effective and the coupling is 
for all practical purposes turned off.
Two-qubit logical gates in this setup can therefore
be controlled by individually tuning the qubits.
Moreover, single-qubit and two-qubit logical operations on different
qubits and pairs of qubits can both be realized simultaneously, 
a requirement to reach presently known thresholds for fault-tolerant
quantum computation \cite{aharonov:96}.

It is interesting to point out that the dispersive QND readout presented in section
\ref{sec:read-out} may be able to determine the state of multiple qubits in a
single shot without the need for additional signal ports. For example, for the
case of two qubits with different detunings, the cavity pull will
take four different values $\pm g_1^2/\Delta_1 \pm
g_2^2/\Delta_2$ allowing single-shot readout of the coupled
system.  This can in principle be extended to $N$ qubits provided
that the range of individual cavity pulls can be made large enough
to distinguish all the combinations. Alternatively, one could read
them out in small groups at the expense of having to electrically
vary the detuning of each group to bring them into strong coupling
with the resonator.

\section{Encoded universality and decoherence-free subspace}
\label{sec:dfs}

Universal quantum computation can also be realized in this
architecture under the encoding $\mathcal{L}=
\{\ket{\uparrow\downarrow},\ket{\downarrow\uparrow}\}$ by
controlling only the qubit's detuning and, therefore, by
turning on and off the interaction term in $H_{2q}$ \cite{lidar:2002c}.

An alternative encoded two-qubit logical operation to
the one suggested in Ref.~\cite{lidar:2002c} can be realized here
by tuning the four qubits forming the pair of encoded qubits
in resonance for a time $t =\pi\Delta/3g^2$.
The resulting effective evolution operator can be written as
$\hat U_{2q}=\exp\left[-i(\pi\Delta/3g^2)\hat\sigma_{x1}\hat\sigma_{x2}\right]$,
where $\hat\sigma_{xi}$ is a Pauli operator acting on
the $i^{\rm th}$ encoded qubit.  Together with encoded
one-qubit operations, $\hat U_{2q}$ is sufficient for universal
quantum computation using the encoding $\mathcal{L}$.

We point out that the subspace $\mathcal{L}$ is a decoherence-free
subspace with respect to global dephasing \cite{kempe:2001} and use of
this encoding will provide some protection against noise.
The application of $\hat U_{2q}$ on the encoded subspace $\mathcal{L}$
however causes temporary leakage out of this protected subspace. 
This is also the case with the approach of Ref.~\cite{lidar:2002c}.  In the present 
situation however, since the Hamiltonian generating $\hat U_{2q}$ commutes with
the generator of global dephasing, this temporary excursion out of the
protected subspace does not induce noise on the encoded qubit.

%%%%%%%%%%%%%%%%%%%%%%%%
% VII: Summary and Conclusions
%%%%%%%%%%%%%%%%%%%%%%%%

\section{Summary and Conclusions}

In summary, we propose that the combination of one-dimensional
superconducting transmission line resonators, which confine their
zero point energy to extremely small volumes, and superconducting
charge qubits, which are electrically controllable qubits with
large electric dipole moments, constitutes an interesting system
to access the strong-coupling regime of cavity quantum
electrodynamics. This combined system is an advantageous
architecture for the coherent control, entanglement, and readout
of quantum bits for quantum computation and communication. Among
the practical benefits of this approach are the ability to
suppress radiative decay of the qubit while still allowing one-bit
operations, a simple and minimally disruptive method for readout
of single and multiple qubits, and the ability to generate tunable
two-qubit entanglement over centimeter-scale distances. We also
note that in the structures described here, the emission or
absorption of a single photon by the qubit is tagged by a sudden
large change in the resonator transmission properties
\cite{smgunpublished} making them potentially useful as single
photon sources and detectors.

\begin{acknowledgments}
We are grateful to David DeMille, Michel Devoret, Clifford Cheung and
Florian Marquardt for useful conversations. We also thank
Andr\'e-Marie Tremblay and the Canadian Foundation for Innovation
for access to computing facilities. This work was supported in
part by the National Security Agency (NSA) and Advanced Research and
Development Activity (ARDA) under Army Research Office (ARO)
contract number DAAD19-02-1-0045, NSF DMR-0196503, NSF DMR-0342157,
the David and Lucile Packard Foundation, the W.M. Keck Foundation and NSERC.
\end{acknowledgments}

%% Appendix
\appendix

\section{Quantization of the 1D transmission line resonator}
\label{appendix:resonator}

A transmission line of length $L$, whose cross section dimension 
is much less then the wavelength of the transmitted signal can be
approximated by a 1-D model.  For relatively low frequencies it is
well described by an infinite series of inductors with each node capacitively
connected to ground, as shown in Fig.~\ref{fig:resonatorandbox}.
Denoting the inductance per unit length $l$ and the capacitance per
unit length $c$, the Lagrangian of the circuit is
\be
\mathcal{L}=\int_{-L/2}^{L/2} dx \left( \frac{l}{2}j^2-\frac{1}{2c}q^2 \right),
\ee
where $j(x,t)$ and $q(x,t)$ are the local current and charge density,
respectively. We have ignored for the moment the two semi-infinite
transmission lines capacitively coupled to the resonator.
Defining the variable $\theta(x,t)$
\be
\theta(x,t)\equiv\int_{-L/2}^x dx'\, q(x',t),
\ee
the Lagrangian can be rewritten as
\be
\mathcal{L}
=\int_{-L/2}^{L/2} dx \left( \frac{l}{2}\dot{\theta}^2-\frac{1}{2c}(\nabla \theta)^2 \right).
\label{eq:lagrangian_theta}
\ee
The corresponding Euler-Lagrange equation is a wave equation with
the speed $v=\sqrt{1/lc}$. Using the boundary conditions
due to charge neutrality
\be
\theta(-L/2,t)=\theta(L/2,t)=0,
\label{eq:boundary_condition}
\ee
we obtain
\begin{eqnarray}
\theta(x,t)
&=&
\sqrt{\frac{2}{L}}
\sum_{k_{\rm o}=1}^{k_{\rm o,cutoff}}\phi_{k_{\rm o}}(t)\cos\frac{k_{\rm o}\pi x}{L}
\nonumber\\
&&+
\sqrt{\frac{2}{L}}
\sum_{k_{\rm e}=2}^{k_{\rm e,cutoff}}\phi_{k_{\rm e}}(t)\sin\frac{k_{\rm e}\pi x}{L},
\label{eq:mode_expansion}
\end{eqnarray}
for odd and even modes, respectively.
For finite length $L$,  the transmission line acts as a resonator
with resonant frequencies $\omega_k=k\pi v/L$.  The cutoff is
determined by the fact that the resonator is not strictly one dimensional.

Using the normal mode expansion \eq{eq:mode_expansion}
in \eq{eq:lagrangian_theta}, one obtains, after spatial integration,
the Lagrangian in the form of a set of harmonic oscillators
\be
\mathcal{L}
= \sum_k 
\frac{l}{2}\dot{\phi_k}^2-\frac{1}{2c}\left(\frac{k\pi}{L}\right)^2\phi_k^2.
\label{eq:lagrangian_oscillators}
\ee

Promoting the variable $\phi_k$ and its canonically conjugated momentum
$\pi_k=l\dot\phi_k$ to conjugate operators and introducing the boson
creation and annihilation operators $a_k^\dag$ and $a_k$ satisfying
$[a_k,a_{k'}^\dag]=\delta_{kk'}$, we obtain the usual relations diagonalizing
the Hamiltonian obtained from the Lagrangian \eq{eq:lagrangian_oscillators}
\begin{eqnarray}
\hat{\phi}_k(t) &=&
\sqrt{\frac{\hbar\omega_k c}{2}}\frac{L}{k\pi}(a_k(t)+a_k^\dag(t))\\
\hat{\pi}_k(t) &=&
-i\sqrt{\frac{\hbar\omega_k l}{2}}(a_k(t)-a_k^\dag(t)).
\end{eqnarray}
From these relations, the voltage on the resonator can be expressed as
\begin{eqnarray}
V(x,t)
&=& \frac{1}{c} \frac{\partial\theta(x,t)}{\partial x}
\\
&=& -\sum_{k_{\rm o}=1}^\infty\sqrt{\frac{\hbar\omega_{k_{\rm o}}}{Lc}}
\sin\left(\frac{k_{\rm o}\pi x}{L}\right)[a_{k_{\rm o}}(t)+a_{k_{\rm o}}^\dag(t)]
\nonumber\\
&& + \sum_{k_{\rm e}=1}^\infty\sqrt{\frac{\hbar\omega_{k_{\rm e}}}{Lc}}
\cos\left(\frac{k_{\rm e}\pi x}{L}\right)[a_{k_{\rm e}}(t)+a_{k_{\rm e}}^\dag(t)].
\nonumber
\end{eqnarray}

In the presence of the two semi-infinite transmission lines coupled
to the resonator, the Lagrangian \eq{eq:lagrangian_theta} and the boundary
conditions \eq{eq:boundary_condition}
are modified to take into account the voltage drop on the coupling capacitors $C_0$.
Assuming no spatial extent for the capacitors $C_0$, the problem is still solvable
analytically. Due to this coupling, the wavefunction can now extend outside
of the central segment which causes a slight red-shift, of order $C_0/Lc$, 
of the cavity resonant frequency.

As shown in Fig.~\ref{fig:resonatorandbox}, we assume the qubit to be fabricated
at the center of the resonator.  As a result, at low temperatures, the qubit is
coupled to the mode $k=2$ of the resonator, which as an anti-node of the
voltage in its center.  The rms voltage between the center conductor and
the ground plane is then ${V}_{\rm rms}^0 = \sqrt{\hbar\omegar / c L}$
with $\omegar  = \omega_2$ and the voltage felt by the qubit is 
$V(0,t) = {V}_{\rm rms}^0 (a_2(t)+a^\dag_2(t))$.  In the main body
of this paper, we work only with this second harmonic and drop the
mode index on the resonator operators.

\section{Quantum non-demolition measurements}
\label{appendix:QND}

Read-out of a qubit can lead to both mixing and dephasing 
\cite{schoelkopf:2003,devoret:2000}.  While dephasing is
unavoidable, mixing of the measured observable
can be eliminated in a QND measurement
by choosing the qubit-measurement apparatus
interaction such that the measured observable is a
constant of motion.  In that situation, the measurement-induced
mixing is rather introduced in the operator conjugate to the operator being
measured.

In the situation of interest in this paper, the operator being 
probed is $\sigma_z$ and, from Eq.~\eq{eq:Jaynes-Cummings-diagonal},
the qubit-measurement apparatus interaction Hamiltonian
is $H_{\rm int} = (g^2/\Delta)\sigma_za^\dag a$, such that $[\sigma_z,H_{\rm int}]=0$.
For $\sigma_z$ to be a constant of motion also requires that
it commutes with the qubit Hamiltonian.  This condition is also satisfied
in Eq.~\eq{eq:Jaynes-Cummings-diagonal}.

That the measured observable is a constant of motion
implies that repeated observations will yield the same result.  
This allows for the measurement result to reach arbitrary large
accuracy by accumulating signal. In practice however, there are always environmental 
dissipation mechanisms acting on the qubit independently of the read-out.  
Even in a QND situation, these will lead to a finite mixing rate $1/T_1$ 
of the qubit in the course of the measurement.  Hence, high fidelity
can only be achieved by a strong measurement completed
in a time $T_{\rm m} \ll T_1$.  This simple point is not as
widely appreciated as it should be.

%% Bibliography
%\bibliographystyle{apsrev}
%\bibliography{ref}

\begin{thebibliography}{43}
\bibitem[{not({\natexlab{b}})}]{note_pull}
\bibinfo{note}{We note that for the case of $Q=10^6$, the cavity pull is a
  remarkable $\pm250$ line widths, but, depending on the non-radiative decay
  rate of the qubit, this may be in the regime $\kappa < \gamma$, making the
  state measurement too slow.}
\end{thebibliography}

\end{document}